\documentclass{article}
           \usepackage{spconf}

           \usepackage{xcolor}
\usepackage{amssymb}
\usepackage{graphicx}

\usepackage{subcaption}
\usepackage{caption} 
\usepackage{stfloats} 
\usepackage{arydshln}

\usepackage{amsmath}
\usepackage{amssymb}

\title{Trigger Sound Suppression for Misophonia}

\newcommand{\squishlist}{\begin{itemize}[itemsep=1pt,parsep=2pt,topsep=3pt,partopsep=0pt,leftmargin=0em, itemindent=1em,labelwidth=1em,labelsep=0.5em]}
\newcommand{\squishend}{\end{itemize}}

\name{Vaishnavi Vidyasagar, 	Jasmine Zhang, Mahima Uliyar,  Seunghyun Oh, Emily Pisani,  Zachary Rosenthal, Shyamnath Gollakota }

\name{\begin{tabular}{@{}c@{}}
Vaishnavi Vidyasagar,$^*$ Jasmine Zhang,$^*$ Mahima Uliyar,$^*$ \\Seunghyun Oh,$^*$ Emily Catherine Gates,$^1$ 
Mark  Zachary Rosenthal,$^1$ Shyamnath Gollakota$^*$
\end{tabular}}

\address{$^*$ University of Washington, $^1$Duke Center for Misophonia and Emotion Regulation}

\begin{document}

\maketitle

\begin{abstract}

Misophonia, a disorder of decreased tolerance to specific sounds, affects
5--20\% of the population, yet sufferers have no good options: therapy
helps a minority, and earplugs or noise cancellation silence everything.
We present a study for neural trigger  sound suppression for
misophonia, selectively removing trigger sounds. We curate a dataset covering the 10 most
common trigger classes. Using streaming dual-path networks operating on 6~ms audio
chunks, we explore both one-hot and multi-hot-conditioned models  that suppress  
1--3 triggers from the acoustic scene. We validate our model outputs in a listening
study with 30 adults with clinically elevated misophonia impairment.  Participants
reported significantly lower distress and arousal, and improved valence,
for suppressed audio.

\end{abstract}

\section{Introduction}
\label{sec:intro}

Misophonia is a disorder characterized by decreased tolerance to specific sounds~\cite{Jakubovski2022, Wu2014, Vitoratou2023}. Individuals experience intense emotional and physiological reactions, including anger, anxiety, disgust, and fight-or-flight responses, to specific ``trigger'' sounds. Common triggers are oral and nasal sounds produced by others, such as chewing, lip smacking, slurping, throat clearing, sniffling, and heavy breathing~\cite{Rosenthal2021DMQ}. Because these sounds are pervasive, misophonia can cause social withdrawal, strained relationships, and impaired  functioning~\cite{Dixon2024}.

 Many individuals  rely on coping strategies with their own costs: earplugs and noise-canceling headphones suppress \textit{all} sound, isolating users from conversation and important auditory cues, while white noise masking~\cite{Jastreboff2015} can cause listening fatigue. Ideally we need selective suppression: remove trigger sounds while not suppressing  the rest of the acoustic scene.

Recent advances in neural target sound extraction (TSE) \cite{Delcroix2022SoundBeam,Veluri2023RealTime} make such selective suppression technically plausible. However, prior systems are primarily evaluated  for general-purpose sound classes like sirens, birds, speech, and music. {Whether they can suppress misophonia triggers remains underexplored}. Success must also be measured not in decibels, but in reduced patient distress. No prior work provides isolated recordings of common misophonia triggers at a scale suitable for training separation models, nor validates suppression with people who have misophonia. 

In this paper, we explore deep learning-based trigger sound suppression for misophonia. Our contributions are: 

\vskip 0.02in\noindent{\bf 1. A misophonia trigger dataset.}
We curate a dataset for   ten  common   trigger classes, i.e.,  chewing, coughing, sneezing, throat clearing, lip smacking, drinking/slurping, heavy breathing, nose sniffling, gum popping, tapping.

\vskip 0.02in\noindent{\bf 2. A streaming, on-device suppression model.} We train models   conditioned on  one-hot and  multi-hot queries that suppress   one to three triggers from the acoustic scene. We use  a streaming variant of a dual-path neural network and 
train 0.496M-parameter models that operate with 
10\,ms algorithmic latency and run in real time on an Orange Pi device.  We also  ablate the STFT parameters to characterize  latency-quality tradeoff down to a fully causal 4\,ms configuration.

\vskip 0.02in\noindent{\bf 3. Validation with misophonic listeners.}
We evaluate model outputs in a listening study with $N=30$ adults with clinically elevated misophonia impairment, using chewing, a  commonly reported primary trigger. Suppressed clips significantly reduced subjective distress and arousal and improved valence relative to raw audio ($p<.001$).\footnote{Code, models and datasets will be open-sourced.}

\section{Related Work}
\label{sec:related}

\textbf{Misophonia and its treatment.} Misophonia is a disorder of decreased tolerance to specific sounds or their associated stimuli~\cite{Swedo2022}.  Representative population studies estimate a prevalence of roughly 5-20\%, depending on the instrument and threshold~\cite{Jakubovski2022, Wu2014, Vitoratou2023}, while a recent nationally representative U.S. study estimated 4.6\% at clinical severity~\cite{Dixon2024}.  A randomized controlled trial  of psychotherapy found clinical improvement in only 37\% of patients~\cite{Jager2020}, and no pharmacological treatment exists. Audiological strategies adapted from tinnitus retraining therapy, including ear-level noise generators and masking~\cite{Jastreboff2015}, offer only partial and non-selective relief.
 We explore a complementary  approach that selectively removes trigger sounds from the acoustic scene in real time.

\noindent\textbf{Target sound extraction.} TSE extracts  a target sound from a mixture using cues such as class labels~\cite{Ochiai2020,Luo2019ConvTasNet,Subakan2021Sepformer,Veluri2023RealTime,Veluri2023SemanticHearing}, enrollment audio~\cite{Delcroix2022SoundBeam}, or text~\cite{Liu2023SeparateAnything}. More recent work enables streaming separation on low-power hardware~\cite{Itani2025TFMLPNet,Oh2026Aurchestra}. While these systems show that real-time, class-conditioned extraction is feasible, they focus on general environmental sounds and have not been evaluated on misophonia triggers. 

\noindent\textbf{Real-time hearable constraints and clinical validation.}
Hearables require low latency, with delays above 20-30~ms being perceptible~\cite{Veluri2023SemanticHearing}. Prior work uses streaming models, low-latency STFTs~\cite{Wang2022LowLatency}, and on-device accelerators~\cite{Itani2025NeuralAids}, but has not evaluated performance on misophonia trigger classes. Clinical studies typically assess audio quality, whereas misophonia requires measuring distress. We address both gaps with per-trigger separation metrics and  distress ratings~\cite{Rosenthal2021DMQ}.
 
\section{Method}
\label{sec:method}
{\bf Problem formulation.} Let $\mathbf{x}[n]\in\mathbb{R}^{2}$ be a binaural scene of sources
$\mathbf{s}_i[n]$ with labels $\ell_i\in\{0,\dots,C\}$, where
$\ell_i\ge1$ indexes a trigger class and $\ell_i=0$ all non-trigger sources.

The user specifies which triggers bother them through a
\emph{multi-hot} query $\mathbf{q}\in\{0,1\}^{C}$. Given $\mathbf{q}$, the scene
decomposes into a trigger component and a residual,
\begin{equation}
\mathbf{t}_{\mathbf{q}}[n]=\!\!\sum_{i\,:\,\ell_i\in\mathcal{Q}}\!\!
\mathbf{s}_i[n],
\qquad
\mathbf{r}_{\mathbf{q}}[n]=\mathbf{x}[n]-\mathbf{t}_{\mathbf{q}}[n].
\label{eq:decomp}
\end{equation}
The desired output is the channel-averaged residual signal, 
i.e., the full scene minus the queried triggers. %

\begin{table*}[t!]
\centering
\footnotesize
\setlength{\tabcolsep}{4pt}
\caption{{Number of audio clips per trigger class in our dataset. }}\vskip -0.15in
\label{tab:dataset_counts}
\begin{tabular}{l|c|c|c|c|c|c|c|c|c|c|c}
\hline
Trigger &
\shortstack{Chewing} &
\shortstack{Cough} &
\shortstack{Drinking} &
\shortstack{Gum pop} &
\shortstack{Heavy breath} &
\shortstack{Lip smack} &
\shortstack{Sneezing} &
\shortstack{Sniffling} &
\shortstack{Tapping} &
\shortstack{Throat clearing} &
Total \\
\hline
FSD50K     &    231   &    -   &  -     &   -  &    -   &  -   &   -  &     -  &    -   &   -    &   231     \\
\hline
CoughVID   &   -    & 1,187 &   -    &   -  &  -     &  -   &   -  & -      &  -     &   -    & 1,187  \\
\hline
ESC-50     &   -    & 40    & 40    &  -   &     -  & -    & 20  &  -     &   -    &  -     & 100    \\
\hline
MATA       &   -    &   -    & 34    &   -  & -      & 21   & 12  & 28    &    -   & -      & 95     \\
\hline
VocalSound &    -   &  -     &   -    &  -   &  -     &   -  & 332 &   -    &     -  & 3,504 & 3,836  \\
\hline
Freesound  & -   &  -     &  -     &  4   &  -     &  -   & 38  & 8     & 149   &   -    & 199    \\
\hline
Deeply     & -   &  -     &  -     &  -   &  -     & 45   &  -  &  -    &  -    &   -    & 45     \\
\hline
YouTube    & 1,191 &     -  & 1,530 & 455 & 1,175 & 885 &  -   & 1,219 & 1,208 &   -    & 7,663  \\
\hline
All        & 1,422 & 1,227 & 1,604 & 459 & 1,175 & 951 & 402 & 1,255 & 1,357 & 3,504 & 13,356 \\
\hline
\end{tabular}
\vskip -0.15in
\end{table*}

\vskip 0.05in\noindent{\bf Misophonia trigger dataset.} Existing corpora cover our  triggers only partially. The AudioSet ontology, whose vocabulary FSD50K inherits,  has no classes for lip smacking, gum popping, or drinking/slurping. Where classes exist, labels are weak (clip-level) and triggers typically co-occur with other sounds. We therefore manually curate a   dataset covering the top ten clinically motivated classes across oral, nasal, and repetitive triggers. Candidate clips are drawn from FSD50K~\cite{fsd50k}, ESC-50~\cite{esc50}, Freesound tags~\cite{freesound}, CoughVID~\cite{coughvid}, MATA~\cite{mata}, VocalSound~\cite{vocalsound}, Deeply Nonverbal Vocalization Dataset~\cite{deeply_nonverbal} and YouTube videos (to be released as metadata), then verified by two human annotators who (i) confirm the class and (ii) ensure the trigger is isolated without background noise. The resulting dataset contains 13,356 clean trigger clips across 10 classes.
 
 For background sources, we combine non-trigger FSD50K and ESC-50 classes (environmental and domestic sounds), TAU Urban Acoustic Scenes 2024 Mobile Evaluation~\cite{tau2024}, and YouTube videos curated to not have the trigger classes. We  include speech and music because they overlap spectrally with oral triggers. The dataset contains 26,742 background clips: 5,142 real recordings and 21,600 TAU urban scene clips from  environments such as restaurants, classrooms, and transportation. We filtered the available TAU clips for ones only 10 seconds long. For both trigger and background sources, we split $80/10/10\%$ into train/val/test with no overlap.

Training mixtures are generated on the fly. {Binaural mixtures are synthesized by convolving each source,  with the head-related impulse responses (HRIRs)   from the  CIPIC HRTF database~\cite{algazi2001cipic}. Each source is independently assigned a direction drawn uniformly from the 1,250 measured HRIR positions.} Each $5$ second  example draws 1-3  trigger events from distinct classes and a  background source. {Each trigger clip is placed at a random position in (or randomly cropped to) the 5 s window,} 
and scaled to a trigger-to-background ratio drawn from $[0,15]$~dB. {If clips are under 5 seconds, the clip is zero-padded to 5s with a random amount of silence in front and the rest at the end.} We focus on this regime, where triggers are at least as loud as the background, because clearly audible triggers are the most disruptive for the target population. Triggers quieter than the background are left to future work.  %
{Evaluation and test mixtures are generated deterministically from fixed per-index seeds, so they are exactly reproducible from the released code and data splits.}

\vskip 0.05in\noindent{\bf Network architecture.}
We use the streaming TF-GridNet~\cite{Wang2023TFGridNet} variant from~\cite{Oh2026Aurchestra}. It takes the two-channel (binaural) mixture $\mathbf{x}\in\mathbb{R}^{2\times N}$ and directly outputs an  estimate of  $r_{\mathbf{q}}[n]$.

The real and imaginary parts of the short-time Fourier transform (STFT) of both channels are stacked and projected to latent channels by a causal convolution with layer normalization. A stack of TF-GridNet blocks follows, each with a residual {spectral} stage (a bidirectional LSTM along frequency) and a residual {temporal} stage (a unidirectional LSTM along time for every frequency bin). A causal transposed convolution then maps the latent channels to two output channels, the real and imaginary parts of the  spectrogram.

Each STFT frame consists of the current chunk of $L_C$ samples as well as $L_B$ lookback  and $L_F$ lookahead samples. Following~\cite{Wang2022LowLatency,Itani2025NeuralAids}, the synthesis window is zero over the first $L_B$ samples and solved for  reconstruction over the remaining $L_C+L_F$, so the algorithmic latency is $(L_C+L_F)/f_s$. We use $L_C/L_B/L_F=6/6/4$\,ms (256 samples at $f_s=16$\,kHz).

The multi-hot query $\mathbf{q}\in\{0,1\}^{C}$ is injected by feature-wise linear modulation (FiLM)~\cite{perez2018film} at the input of every TF-GridNet block: two linear layers map $\mathbf{q}$ to a per-channel scale and shift applied to the latent channels.

We train two classes of models:  (i) a \emph{single-trigger multi-class} model over all $C$ trigger classes with $|\mathcal{Q}|=1$ at both train and test; and (ii) the \emph{multi-trigger multi-hot queried} model trained with $|\mathcal{Q}|\geq 1$ as described above. We also train and compare on the same datasets with  Waveformer~\cite{Veluri2023RealTime}, which is a  time-domain based target sound extraction model.

\begin{table}[t!]
\centering
\caption{Performance of  a single-trigger multi-class  model trained across all 10 trigger classes, evaluated with single-trigger mixtures and a single output channel.}
\vskip -0.1in
\label{tab:ours_si_sdri_snri}
\begin{tabular}{lcc}
\hline
{Trigger Class} & {SI-SNRi (dB)} & {SNRi (dB)} \\
\hline
Chewing & $11.88_{\pm 4.36}$ & $13.18_{\pm 4.26}$ \\
Cough & $15.79_{\pm 5.47}$ & $15.91_{\pm 5.30}$ \\
Drinking & $10.73_{\pm 3.52}$ & $11.74_{\pm 3.48}$ \\
Gum popping & $13.29_{\pm 4.07}$ & $13.85_{\pm 3.96}$ \\
Heavy Breathing & $12.48_{\pm 3.79}$ & $13.31_{\pm 3.81}$ \\
Lip smacking & $14.35_{\pm 5.27}$ & $15.07_{\pm 5.05}$ \\
Sneezing & $16.47_{\pm 5.48}$ & $16.74_{\pm 5.27}$ \\
Sniffling & $16.91_{\pm 5.21}$ & $17.08_{\pm 5.15}$ \\
Tapping & $13.27_{\pm 3.83}$ & $14.09_{\pm 3.75}$ \\
Throat Clearing & $14.81_{\pm 4.71}$ & $15.09_{\pm 4.63}$ \\
\hline
\end{tabular}
\vskip -0.1in
\end{table}

\section{Experiments and Results}

 \noindent{\bf Hyperparameters.}  We use the OrangePi configuration in~\cite{Oh2026Aurchestra} with six GridNet layers and a 32-dimensional latent space, giving an algorithmic latency of 10\,ms (6\,ms chunk plus 4\,ms lookahead). Models are trained in the same chunked mode used at inference so that train and test state handling match. We used AdamW with an initial learning rate of $1\times10^{-4}$, no weight decay, a batch size of 2 per rank, and gradient clipping with the norm set to 1. 
 We halved the learning rate if validation SNRi did not improve over five epochs, and selected the best-performing checkpoint on the validation set.
 The multi-class single-trigger  models were trained for 100 epochs with 30{,}000 samples per epoch (3,000 samples per  trigger class), so that every class is seen within each epoch. The multi-trigger model was not trained from scratch: we initialized it from the best single-trigger checkpoint and fine-tuned for 50 further epochs under the same settings. {The number of trigger sounds per mixture depends on the model type: single-trigger multi-class models see exactly one trigger per mixture, while the multi-trigger model sees 1, 2, or 3 triggers, drawn uniformly at random with equal probability. For a given mixture, trigger classes are drawn uniformly at random from the 10 classes without repetition, so all class combinations are equally likely and no class is over- or under-represented by its clip count; for each selected class, a clip is drawn uniformly at random from that class's training clips, and a random 5-second segment is extracted, resampled if it is judged near-silent.} %

\begin{table}[t!]
\centering
\setlength{\tabcolsep}{2pt}
\small
\caption{Multi-trigger  suppression model with 0.496M parameters, compared to 2.020M for  Waveformer baseline.}
\vskip -0.1in
\label{tab:multi_target}
\begin{tabular}{llccc}
\hline
 & & \multicolumn{3}{c}{\# selected trigger classes} \\
Model & Metric & 1 & 2 & 3 \\
\hline
Ours & SI-SNRi & $19.86_{\pm 5.82}$ & $18.17_{\pm 5.23}$ & $16.72_{\pm 5.10}$ \\
     & SNRi    & $19.99_{\pm 5.47}$ & $18.69_{\pm 4.82}$ & $17.78_{\pm 4.43}$ \\
\hline
Waveformer & SI-SNRi & $12.92_{\pm 5.06}$ & $12.30_{\pm 4.17}$ & $11.90_{\pm 3.88}$ \\
           & SNRi    & $13.67_{\pm 4.87}$ & $13.76_{\pm 4.07}$ & $14.07_{\pm 3.93}$ \\
\hline
\end{tabular}
\vskip -0.1in
\end{table}

\vskip 0.05in\noindent{\bf Loss function.} All  models are trained with a composite loss  between the ground truth residue ${s}$ and its prediction $\hat{s}$ : 
\begin{equation}
\mathcal{L} =
\mathcal{L}_{\mathrm{MRSTFT}}(\hat{s}, s)
+ \lambda_{1}\,\lVert \hat{s} - s \rVert_1
- \lambda_{\mathrm{snr}}\,\mathrm{SNR}(\hat{s}, s)
\label{eq:loss}
\end{equation}
where $\mathcal{L}_{\mathrm{MRSTFT}}$ is a multi-resolution STFT loss~\cite{Oh2026Aurchestra} and  $\lVert\cdot\rVert_1$ is the $L_1$ loss. We set
$\lambda_1 = 10$ and $\lambda_{\mathrm{snr}} = 0.1$.

\begin{table}[t!]
\centering
\setlength{\tabcolsep}{3pt}
\caption{{Ablation study over  STFT parameters for the multi-trigger  model, evaluated with a single trigger queried.}}
\vskip -0.1in
\label{tab:latency}
\begin{tabular}{lccc}
\hline
 ($L_C/L_B/L_F$, ms) & Latency & SNRi (dB) & SI-SNRi (dB) \\
\hline
6/6/4 & 10\,ms & $19.99_{\pm 5.47}$ & $19.86_{\pm 5.82}$ \\
8/4/4 & 12\,ms & $17.08_{\pm 5.02}$ & $16.76_{\pm 5.25}$ \\
4/12/0  & 4\,ms & $13.38_{\pm 4.84}$ & $12.55_{\pm 5.08}$ \\
\hline
\end{tabular}
\vskip -0.1in
\end{table}

\vskip 0.08in\noindent{\bf Results.} Table~\ref{tab:ours_si_sdri_snri} reports the results for our single-trigger multi-class  model. The table shows both the per-class residual signal SI-SNRi and SNRi on the
held-out test split over 3,000 test mixtures per trigger class.  {Impulsive,
spectrally distinct triggers such as Sniffling (16.91\,dB) and Sneezing
(16.47\,dB) are separated most reliably, while Drinking (10.73\,dB) and Chewing
(11.88\,dB) are the hardest, as their broadband, low-energy transients overlap
more heavily with background speech and ambient noise.} The gap between residual signal SNRi and
SI-SNRi is under {1.30\,dB} for every class, indicating the residual  signals are
close to correctly scaled rather than relying on a global gain correction.

Table~\ref{tab:multi_target} shows results for the multi-trigger suppression model on test mixtures with multiple triggers. Our model outperforms Waveformer~\cite{Veluri2023RealTime} across all three trigger-count conditions. Its SI-SNR and SNR improvements decrease with more concurrent trigger suppression, since the model has the same parameter count as the single-trigger model but must suppress arbitrary combinations of 1--3 triggers. %

{The ablation study in Table~\ref{tab:latency} examines how the STFT window budget affects performance for the multi-trigger  suppression model.} We retrain the model by changing the current chunk size $L_C$, lookback $L_B$ and lookahead $L_F$, while keeping their sum constant at 16~ms so the STFT length remains fixed. Removing lookahead entirely (4/12/0) yields a fully causal model at $4$\,ms latency, but costs $3.70$\,dB SNRi ($4.21$\,dB SI-SNRi) relative to  8/4/4 window. %

The table reports the algorithmic latency. The specific processing time however depends on the target hardware  running the model. Our network processes each 6~ms chunk in 5.22\,ms on average and 5.41\,ms at the 95th percentile on an Orange Pi 5B (Arm Cortex-A76 at 2.4\,GHz, ONNX).

\begin{figure}[t!]
\centering
\includegraphics[width=\linewidth]{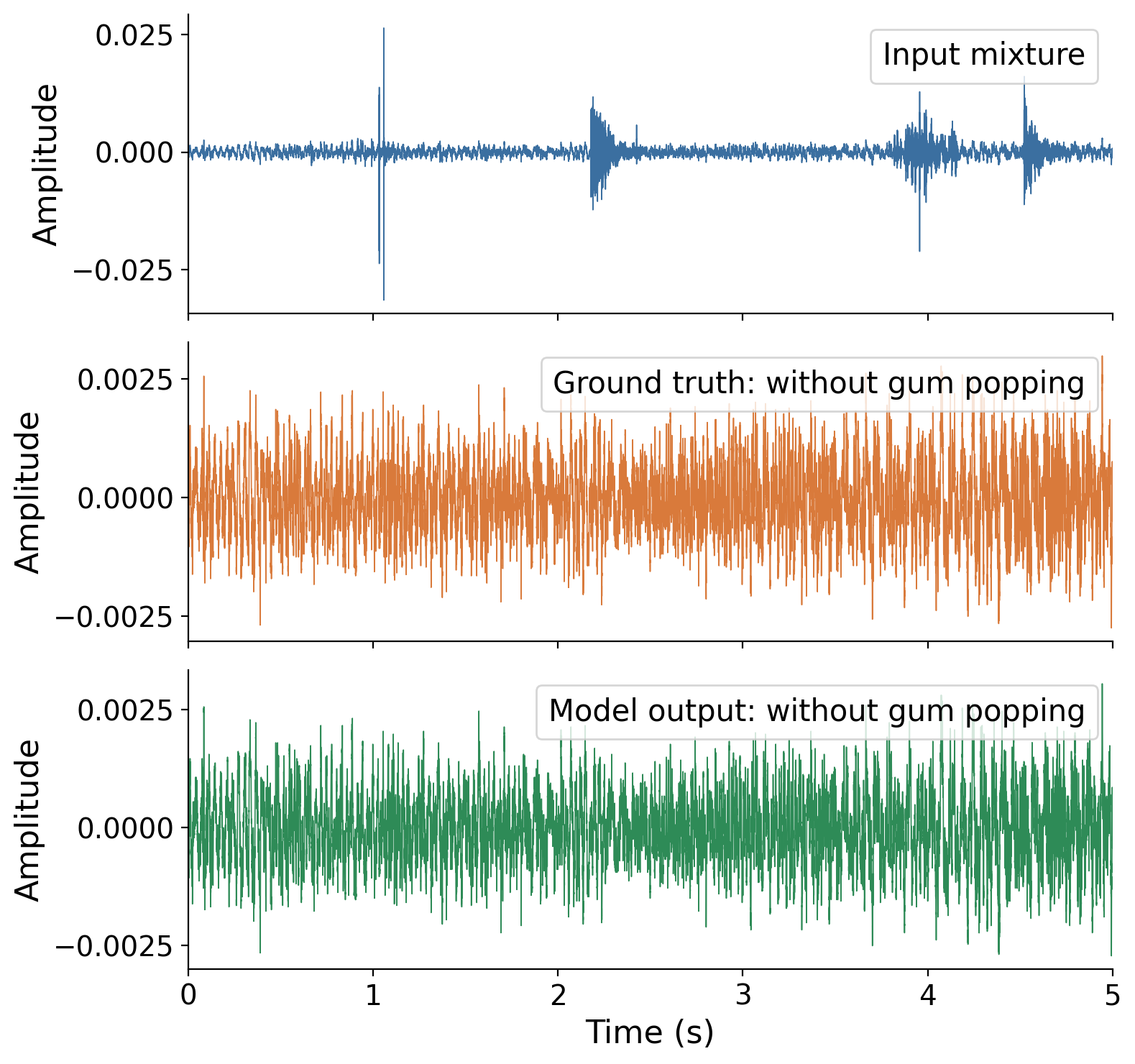}
\vskip -0.15in
\caption{Time-domain waveforms for gum popping suppression: input mixture, ground truth, and model output.}
\vskip -0.15in
\label{fig:gumpop}
\end{figure}

In {Fig.~\ref{fig:gumpop},} we qualitatively show an average example of
single-trigger suppression. The output
closely tracks the temporal envelope and fine structure of the ground truth,  indicating that
the model recovers the full residual signal. The
extracted signals are also correctly scaled, so the model preserves the 
amplitude of the non-trigger sounds in the mixture.
 
\subsection{Listening Study with Misophonic Individuals}
\label{sec:userstudy}
We  ran an online  IRB-approved listening study with $N=30$ adults who reported chewing, a  commonly reported primary trigger, as a bothersome trigger. Participants were recruited via Duke Misophonia mailing lists and social media. Participants had no bilateral hearing loss or profound vision loss.  A Duke Misophonia Questionnaire (DMQ) Impairment score of 14 or higher was required as our inclusive criteria, so all participants met criteria for moderate or higher impairment~\cite{Rosenthal2021DMQ}. Participants scored a mean DMQ Impairment of 31.3 (SD 6.0, median 32) out of a possible 48, indicating a sample skewed toward more significant symptoms. %

Participants were presented with a series of ten random audio samples, each 5 seconds in length. These corresponded to different source clips from our test set, each presented in both an original version and the model output in individually randomized order. Each audio sample contained a mixture of everyday background sounds together with a chewing trigger sound. Both conditions were presented in mono, generated by the multi-trigger model, and participants were not informed which clips had been through the model.  Participants rated their subjective distress (1-10 scale) immediately after each clip, along with emotional arousal and valence (Self-Assessment Manikin scales). A relaxation/reset step was offered between clips to reduce carryover distress.

Table~\ref{tab:userstudy} shows that the model output reduced distress and arousal and improved valence relative to the original clips, with large effect sizes on all three measures. Distress fell by $2.37$ points ($t(29)=5.30$, $p<.001$, $d=0.97$) and arousal by $2.13$ points ($t(29)=5.62$, $p<.001$, $d=1.03$), while valence rose by $1.59$ points ($t(29)=-6.03$, $p<.001$, $d=-1.10$). The three measures move consistently and in the expected directions, indicating that suppression shifts the broader affective response. 
Post-suppression distress remained above floor, likely because,  triggers are idiosyncratic and extend beyond our ten classes. As noted by multiple  participants, background scenes (e.g., restaurants)   contain individually triggering sounds such as cutlery or typing. Since these sounds were present in both conditions, the observed reduction reflects the effect of removing chewing alone.  This is also confirmed  by two non-participating adults who did not report hearing chewing in the model outputs.

Interest in our technology was high (8.73/10), as was the perceived impact of losing access (8.63/10). Expected  benefit was consistently near 8/10 across academic/work (8.10), home (7.93), social (8.57), and general (8.03) domains.

\begin{table}[t!]
\centering
\caption{Affective response to raw audio vs.\ model output (30 participants). Lower is better for distress and arousal; higher better for valence. All paired $t$-tests $p<.001$.}
\vskip -0.15in
\label{tab:userstudy}
\footnotesize
\begin{tabular}{lccc}
\hline
Measure & Original & Output & $\Delta$ [95\% CI] \\
\hline
Distress (1--10) & $7.68 \pm 1.60$ & $5.31 \pm 2.48$ & $-2.37$ [1.45, 3.28] \\
Arousal (1--9)   & $7.02 \pm 1.42$      & $4.89 \pm 2.19$      & $-2.13$ [1.36, 2.91] \\
Valence (1--9)   & $4.42 \pm 2.75$      & $6.01 \pm 1.65$      & $+1.59$ [1.05, 2.13] \\
\hline
\end{tabular}
\vskip -0.15in
\end{table}

\section{Conclusion}
\label{sec:conclusion}
We address a practical problem for people with misophonia: suppressing the
trigger sounds that distress them. %

\noindent\textbf{Discussion.} We note that although active noise cancellation (ANC) itself requires sub-millisecond
processing, recent work~\cite{Veluri2023SemanticHearing, doi:10.1177/23312165241260029} shows that
deep-learning-based hearables can operate within a 10\,ms algorithmic latency by running ANC
to suppress the full acoustic scene and then reinjecting the processed sounds of
interest into the ear canal. Our latency therefore falls within the range this
architecture allows. %

\noindent\textbf{Limitations and future work.} Our dataset was curated using online audio sources, many of which are recorded close to the microphone. Augmenting it with in-situ recordings in  reverberant rooms, captured on hearable microphones, would reduce the gap between our simulated mixtures and real use. Expanding coverage to more triggers is also needed. 
Deploying on the AI accelerators found in hearable form factors will require
further optimization along the lines of~\cite{Itani2025TFMLPNet}. Future work is also required for training models to output binaural signals and perform end-to-end in-person evaluation. %

\section{Acknowledgements}

This work was supported by a grant from the Misophonia Foundation. The authors were also supported by a Moore Foundation fellowship.

{\footnotesize
\bibliographystyle{IEEEtran}
\bibliography{paper}

\begin{thebibliography}{10}
\providecommand{\url}[1]{#1}
\csname url@samestyle\endcsname
\providecommand{\newblock}{\relax}
\providecommand{\bibinfo}[2]{#2}
\providecommand{\BIBentrySTDinterwordspacing}{\spaceskip=0pt\relax}
\providecommand{\BIBentryALTinterwordstretchfactor}{4}
\providecommand{\BIBentryALTinterwordspacing}{\spaceskip=\fontdimen2\font plus
\BIBentryALTinterwordstretchfactor\fontdimen3\font minus
  \fontdimen4\font\relax}
\providecommand{\BIBforeignlanguage}[2]{{%
\expandafter\ifx\csname l@#1\endcsname\relax
\typeout{** WARNING: IEEEtran.bst: No hyphenation pattern has been}%
\typeout{** loaded for the language `#1'. Using the pattern for}%
\typeout{** the default language instead.}%
\else
\language=\csname l@#1\endcsname
\fi
#2}}
\providecommand{\BIBdecl}{\relax}
\BIBdecl

\bibitem{Jakubovski2022}
E.~Jakubovski, A.~M{\"u}ller, H.~Kley, M.~de~Zwaan, and K.~M{\"u}ller-Vahl,
  ``Prevalence and clinical correlates of misophonia symptoms in the general
  population of {Germany},'' \emph{Frontiers in Psychiatry}, vol.~13, p.
  1012424, 2022.

\bibitem{Wu2014}
M.~S. Wu, A.~B. Lewin, T.~K. Murphy, and E.~A. Storch, ``Misophonia: Incidence,
  phenomenology, and clinical correlates in an undergraduate student sample,''
  \emph{Journal of Clinical Psychology}, 2014.

\bibitem{Vitoratou2023}
S.~Vitoratou, C.~Hayes, N.~Uglik-Marucha, O.~Pearson, T.~Graham, and
  J.~Gregory, ``Misophonia in the {UK}: Prevalence and norms from the {S-Five}
  in a {UK} representative sample,'' \emph{PLoS ONE}, 2023.

\bibitem{Rosenthal2021DMQ}
M.~Z. Rosenthal, D.~Anand, C.~Cassiello-Robbins, Z.~J. Williams, R.~E. Guetta,
  J.~Trumbull, and L.~Kelley, ``Development and initial validation of the {Duke
  Misophonia Questionnaire},'' \emph{Frontiers in Psychology}, vol.~12, p.
  709928, 2021.

\bibitem{Dixon2024}
L.~J. Dixon, M.~J. Schadegg, H.~L. Clark, C.~J. Sevier, and S.~M. Witcraft,
  ``Prevalence, phenomenology, and impact of misophonia in a nationally
  representative sample of {U.S.} adults,'' \emph{Journal of Psychopathology
  and Clinical Science}, 2024.

\bibitem{Jastreboff2015}
P.~J. Jastreboff and M.~M. Jastreboff, ``Decreased sound tolerance:
  Hyperacusis, misophonia, diplacousis, and polyacousis,'' in \emph{Handbook of
  Clinical Neurology}.\hskip 1em plus 0.5em minus 0.4em\relax Elsevier, 2015.

\bibitem{Delcroix2022SoundBeam}
M.~Delcroix, J.~B. Vázquez, T.~Ochiai, K.~Kinoshita, Y.~Ohishi, and S.~Araki,
  ``Soundbeam: Target sound extraction conditioned on sound-class labels and
  enrollment clues for increased performance and continuous learning,''
  \emph{IEEE/ACM Transactions on Audio, Speech, and Language Processing},
  vol.~31, pp. 121--136, 2023.

\bibitem{Veluri2023RealTime}
B.~Veluri, J.~Chan, M.~Itani, T.~Chen, T.~Yoshioka, and S.~Gollakota,
  ``Real-time target sound extraction,'' in \emph{ICASSP 2023 - 2023 IEEE
  International Conference on Acoustics, Speech and Signal Processing
  (ICASSP)}, 2023, pp. 1--5.

\bibitem{Swedo2022}
S.~E. Swedo, D.~M. Baguley, D.~Denys, L.~J. Dixon, M.~Erfanian, A.~Fioretti,
  P.~J. Jastreboff, S.~Kumar, M.~Z. Rosenthal, R.~Rouw, D.~Schiller, J.~Simner,
  E.~A. Storch, S.~Taylor, K.~R.~V. Werff, C.~M. Altimus, and S.~M. Raver,
  ``Consensus definition of misophonia: A {Delphi} study,'' \emph{Frontiers in
  Neuroscience}, vol.~16, p. 841816, 2022.

\bibitem{Jager2020}
I.~J. Jager, N.~C. Vulink, I.~O. Bergfeld, A.~J. van Loon, and D.~A. Denys,
  ``Cognitive behavioral therapy for misophonia: A randomized clinical trial,''
  \emph{Depression and Anxiety}, 2020.

\bibitem{Ochiai2020}
T.~Ochiai, M.~Delcroix, Y.~Koizumi, H.~Ito, K.~Kinoshita, and S.~Araki,
  ``{Listen to What You Want: Neural Network-Based Universal Sound Selector},''
  in \emph{{Interspeech 2020}}, 2020, pp. 1441--1445.

\bibitem{Luo2019ConvTasNet}
\BIBentryALTinterwordspacing
Y.~Luo and N.~Mesgarani, ``Conv-tasnet: Surpassing ideal time–frequency
  magnitude masking for speech separation,'' \emph{IEEE/ACM Trans. Audio,
  Speech and Lang. Proc.}, vol.~27, no.~8, p. 1256–1266, Aug. 2019. [Online].
  Available: \url{https://doi.org/10.1109/TASLP.2019.2915167}
\BIBentrySTDinterwordspacing

\bibitem{Subakan2021Sepformer}
C.~Subakan, M.~Ravanelli, S.~Cornell, M.~Bronzi, and J.~Zhong, ``Attention is
  all you need in speech separation,'' in \emph{ICASSP 2021 - 2021 IEEE
  International Conference on Acoustics, Speech and Signal Processing
  (ICASSP)}, 2021, pp. 21--25.

\bibitem{Veluri2023SemanticHearing}
B.~Veluri, M.~Itani, J.~Chan, T.~Yoshioka, and S.~Gollakota, ``Semantic
  hearing: Programming acoustic scenes with binaural hearables,'' in
  \emph{Proc. ACM Symposium on User Interface Software and Technology (UIST)},
  2023, pp. 1--15.

\bibitem{Liu2023SeparateAnything}
X.~Liu, Q.~Kong, Y.~Zhao, H.~Liu, Y.~Yuan, Y.~Liu, R.~Xia, Y.~Wang, M.~D.
  Plumbley, and W.~Wang, ``Separate anything you describe,'' \emph{IEEE
  Transactions on Audio, Speech and Language Processing}, vol.~33, pp.
  458--471, 2025.

\bibitem{Itani2025TFMLPNet}
M.~Itani, T.~Chen, and S.~Gollakota, ``{TF-MLPNet}: Tiny real-time neural
  speech separation,'' in \emph{Proc. Clarity Challenge, Interspeech}, 2025.

\bibitem{Oh2026Aurchestra}
S.~Oh, M.~Itani, A.~Gauri, and S.~Gollakota, ``Fine-grained soundscape control
  for augmented hearing,'' \emph{Proceedings of the 24th Annual International
  Conference on Mobile Systems, Applications and Services}, p. 371–391, 2026.

\bibitem{Wang2022LowLatency}
Z.-Q. Wang, G.~Wichern, S.~Watanabe, and J.~Le~Roux, ``Stft-domain neural
  speech enhancement with very low algorithmic latency,'' \emph{IEEE/ACM
  Transactions on Audio, Speech, and Language Processing}, vol.~31, pp.
  397--410, 2023.

\bibitem{Itani2025NeuralAids}
M.~Itani, T.~Chen, A.~Raghavan, G.~Kohlberg, and S.~Gollakota, ``Wireless
  hearables with programmable speech {AI} accelerators,'' in \emph{Proc. ACM
  International Conference on Mobile Computing and Networking (MobiCom)}, 2025.

\bibitem{fsd50k}
E.~Fonseca, X.~Favory, J.~Pons, F.~Font, and X.~Serra, ``{FSD50K}: An open
  dataset of human-labeled sound events,'' \emph{IEEE/ACM Transactions on
  Audio, Speech, and Language Processing}, vol.~30, pp. 829--852, 2022.

\bibitem{esc50}
K.~J. Piczak, ``{ESC}: Dataset for environmental sound classification,'' in
  \emph{Proc. 23rd ACM Int. Conf. Multimedia (MM '15)}, 2015, pp. 1015--1018.

\bibitem{freesound}
F.~Font, G.~Roma, and X.~Serra, ``Freesound technical demo,'' in \emph{Proc.
  21st ACM Int. Conf. Multimedia (MM '13)}, 2013, pp. 411--412.

\bibitem{coughvid}
L.~Orlandic, T.~Teijeiro, and D.~Atienza, ``The {COUGHVID} crowdsourcing
  dataset, a corpus for the study of large-scale cough analysis algorithms,''
  \emph{Scientific Data}, vol.~8, no.~1, p. 156, 2021.

\bibitem{mata}
S.~V. Shinkareva, ``The misophonia audiovisual trigger archive ({MATA}),''
  \emph{bioRxiv}, 2025, preprint. Dataset: \url{https://osf.io/e7qbm/}, DOI
  10.17605/OSF.IO/E7QBM.

\bibitem{vocalsound}
Y.~Gong, J.~Yu, and J.~Glass, ``{VocalSound}: A dataset for improving human
  vocal sounds recognition,'' in \emph{ICASSP 2022 -- IEEE Int. Conf.
  Acoustics, Speech and Signal Processing}, 2022, pp. 151--155.

\bibitem{deeply_nonverbal}
\BIBentryALTinterwordspacing
D.~Inc., ``{Deeply Nonverbal Vocalization dataset},'' 2021. [Online].
  Available: \url{https://github.com/deeplyinc/Nonverbal-Vocalization-Dataset}
\BIBentrySTDinterwordspacing

\bibitem{tau2024}
\BIBentryALTinterwordspacing
T.~Heittola, A.~Mesaros, and T.~Virtanen, ``{TAU Urban Acoustic Scenes 2024
  Mobile, Evaluation dataset},'' 2024. [Online]. Available:
  \url{https://doi.org/10.5281/zenodo.11366913}
\BIBentrySTDinterwordspacing

\bibitem{algazi2001cipic}
V.~R. Algazi, R.~O. Duda, D.~M. Thompson, and C.~Avendano, ``The {CIPIC} {HRTF}
  database,'' in \emph{Proc. IEEE Workshop on Applications of Signal Processing
  to Audio and Acoustics (WASPAA)}, New Paltz, NY, USA, Oct. 2001, pp. 99--102.

\bibitem{Wang2023TFGridNet}
Z.-Q. Wang, S.~Cornell, S.~Choi, Y.~Lee, B.-Y. Kim, and S.~Watanabe,
  ``Tf-gridnet: Making time-frequency domain models great again for monaural
  speaker separation,'' in \emph{ICASSP 2023 - 2023 IEEE International
  Conference on Acoustics, Speech and Signal Processing (ICASSP)}, 2023, pp.
  1--5.

\bibitem{perez2018film}
E.~Perez, F.~Strub, H.~De~Vries, V.~Dumoulin, and A.~Courville, ``Film: Visual
  reasoning with a general conditioning layer,'' in \emph{AAAI}, vol.~32,
  no.~1, 2018.

\bibitem{doi:10.1177/23312165241260029}
\BIBentryALTinterwordspacing
A.~T. Sabin, D.~McElhone, D.~Gauger, and B.~Rabinowitz, ``Modeling the
  intelligibility benefit of active noise cancelation in hearing devices that
  improve signal-to-noise ratio,'' \emph{Trends in Hearing}, vol.~28, p.
  23312165241260029, 2024, pMID: 38831646. [Online]. Available:
  \url{https://doi.org/10.1177/23312165241260029}
\BIBentrySTDinterwordspacing

\end{thebibliography}
}
\end{document}